\documentclass[10pt, conference, twocolumn]{IEEEtran}
\usepackage{mdwmath}
\usepackage{amsmath}
\usepackage{amssymb}
\usepackage{cite}
\usepackage{epsfig}
\newtheorem{rem}{Remark}
\newtheorem{theo}{Theorem}

\newtheorem{lem}{Lemma}

\newcommand{\rank}{\text{rank}}

\newcommand{\diag}{\text{diag}}

\newcommand{\eqdef}{\stackrel{\triangle}{=}}

\usepackage{graphicx}
\usepackage{graphics}
\usepackage{color}
\usepackage{psfrag}
\usepackage{pstricks, pst-node, pst-tree}
\usepackage{pstricks-add}

\begin{document}
\author{Dimitris S. Papailiopoulos and Alexandros G. Dimakis\\
Department of Electrical Engineering\\
University of Southern California\\
Los Angeles, CA 90089\\
Email:\texttt{\{papailio, dimakis\}@usc.edu}}

\title{Distributed Storage Codes through Hadamard Designs}
\maketitle

\begin{abstract}
In distributed storage systems that employ erasure coding, the issue of minimizing the total {\it repair bandwidth} required to exactly regenerate a storage node after a failure arises.
This repair bandwidth depends on the structure of the storage code and the repair strategies used to restore the lost data.
Minimizing it requires that undesired data during a repair align in the smallest possible spaces, using the concept of interference alignment (IA).
Here, a points-on-a-lattice representation of the symbol extension IA of Cadambe {\it et al.} provides cues to perfect IA instances which we combine with fundamental properties of Hadamard matrices to construct a new storage code with favorable repair properties.
Specifically, we build an explicit $(k+2,k)$ storage code over $\mathbb{GF}(3)$, whose single systematic node failures can be repaired with bandwidth that matches exactly the theoretical minimum.
Moreover, the repair of single parity node failures generates at most the same repair bandwidth as any systematic node failure.
Our code can tolerate any single node failure and any pair of failures that involves at most one systematic failure.
\end{abstract}

\section{Introduction}
The demand for large scale data storage has increased significantly in recent years with applications demanding seamless storage, access, and security for massive amounts of data. 
When the deployed nodes of a storage network are individually unreliable, as is the case in modern data centers, or peer-to-peer networks, redundancy through erasure coding can be introduced to offer reliability against node failures. 
However, increased reliability does not come for free:  the encoded representation needs to be maintained posterior to node erasures.
To maintain the same redundancy when a storage node leaves the system, a new node has to join the array, access some existing nodes, and regenerate the contents of the departed node.
This problem is known as the {\it Code Repair Problem} \cite{DimakisGWWR:08}, \cite{storagewiki}.

The interest in the code repair problem, and specifically in designing repair optimal $(n,k)$ erasure codes, stems from the fact that there exists a fundamental minimum repair bandwidth needed to regenerate a lost node that is substantially less than the size of the encoded data object.
MDS erasure storage codes have generated particular interest since they offer maximum reliability for a given storage capacity; such an example is the EvenOdd construction \cite{evenodd}.
However, most practical solutions for storage use existing off-the-shelf erasure codes that are repair inefficient: a  single node repair generates network traffic equal to the size of the {\it entire} stored information.

Designing repair optimal MDS codes, i.e., ones achieving the minimum repair bandwidth bound that was derived in \cite{DimakisGWWR:08}, seems to be challenging especially for high rates $\frac{k}{n}\ge\frac{1}{2}$.
Recent works by Cadambe {\it et al.} \cite{CadambeWinc} and Suh {\it et al.} \cite{SuhCodes} used the symbol extension IA technique of Cadambe {\it et al.} \cite{CadambeJ:08} to establish the existence, for all $n$, $k$, of asymptotically optimal MDS storage codes, that come arbitrarily close to the theoretic minimum repair bandwidth. 
However, these asymptotic schemes are impractical due to the arbitrarily large file size and field size that they require.
Explicit and practical designs for optimal MDS storage codes are constructed roughly for rates $\frac{k}{n}\le\frac{1}{2}$ \cite{WuD:09}-\cite{Wu:09c}, \cite{RashmiProduct}, and most of them are based upon the concept of interference alignment.
Interestingly, as of now no explicit MDS storage code constructions exist with optimal repair properties for the high data rate regime.\footnote{During the submission of this manuscript, two independent works appeared that constructed MDS codes of arbitrary rate that can optimally repair their systematic nodes, see \cite{Tamo}, \cite{PermCodes}.}

{\bf Our Contribution}: In this work we introduce a new high-rate, explicit, $(k+2,k)$ storage code over $\mathbb{GF}(3)$.
Our storage code exploits fundamental properties of Hadamard designs and perfect IA instances pronounced by the use of a lattice representation for the symbol extension IA of Cadambe {\it et al.} \cite{CadambeJ:08}.
This representation gives hints for coding structures that allow {\it exact} instead of asymptotic alignment.
Our code exploits these structures and achieves perfect IA without requiring the file size or field size to scale to infinity.
Any single systematic node failure can be repaired with bandwidth matching the theoretic minimum and any single parity node failure generates (at most) the same repair bandwidth as any systematic node repair.
Our code has two parities but cannot tolerate any two failures: the form presented here can tolerate any single failure and any pair of failures that involves at most one systematic node failure\footnote{Our latest work expands Hadamard designs to construct $2$-parity MDS codes that can optimally repair any systematic or parity node failure and $m$-parity MDS codes that can optimally repair any systematic node failure \cite{PVD}.}.
Here, in contrast to MDS codes, slightly more than $k$, that is, $k\left(1+\frac{1}{2k}\right)$, encoded pieces are required to reconstruct the file object.


\section{Distributed Storage Codes with $2$ Parity Nodes}
In this section, we consider the code repair problem for storage codes with $2$ parity nodes. 
Let a file of size $M=kN$ denoted by the vector ${\bf f}\in\mathbb{F}^{kN}$ be partitioned in $k$ parts ${\bf f}=\left[{\bf f}^T_1\ldots{\bf f}^T_k\right]^T$, each of size $N$.\footnote{$\mathbb{F}$ denotes the finite field over which all operations are performed.}
We wish to store this file with rate $\frac{k}{k+2}$ across $k$ systematic and $2$ parity storage units each having storage capacity $\frac{M}{k}=N$.
To achieve this level of redundancy, the file is encoded using a $(k+2,k)$ distributed storage code.
The structure of the storage array is given in Fig. 1, where ${\bf A}_i$ and ${\bf B}_i$ are $N\times N$ matrices of coding coefficients used by the parity nodes $a$ and $b$, respectively,  to ``mix'' the contents of the $i$th file piece ${\bf f}_{i}$.
Observe that the code is in systematic form: $k$ nodes store the $k$ parts of the file and each of the $2$ parity nodes stores a linear combination of the $k$ file pieces.
\begin{figure}[t]
\begin{align}
&\begin{array}{|c|c|}
\hline
\text{systematic node} & \text{systematic data}\\
\hline
1&{\bf f}_1\\
\hline
\vdots&\vdots\\
\hline
k&{\bf f}_k\\
\hline
\text{parity node} & \text{parity data}\\
\hline
a&{\bf A}_1^T{\bf f}_1+\ldots+{\bf A}_k^T{\bf f}_k\\
\hline
b&{\bf B}_1^T{\bf f}_1+\ldots+{\bf B}_k^T{\bf f}_k\\
\hline
\end{array}\nonumber
\end{align}
\caption{\textsc{A $(k+2,k)$ Coded Storage Array.}}
\end{figure}

To maintain the same level of redundancy when a node fails or leaves the system, the code repair process has to take place to exactly restore the lost data in a {\it newcomer} storage component.
Let for example a systematic node $i\in\{1,\ldots,k\}$ fail.
Then, a newcomer joins the storage network, connects to the remaining $k+1$ nodes, and has to download sufficient data to reconstruct ${\bf f}_i$.
Observe that the missing piece ${\bf f}_i$ exists as a term of a linear combination {\it only} at each parity node, as seen in Fig. 1.
To regenerate it, the newcomer has to download from the parity nodes at least the size of what was lost, i.e., $N$ linearly independent data elements.
The downloaded contents from the parity nodes can be represented as a stack of $N$ equations
{\small
\begin{align}
\hspace{-0.1cm}\left[
\begin{array}{c}
{\bf p}_i^{(a)}\\
{\bf p}_i^{(b)}
\end{array}
\right]\hspace{-0.1cm}&\eqdef\hspace{-0.1cm}\underbrace{\left[
\begin{array}{@{}c@{}}
\left({\bf A}_{i}{\bf V}^{(a)}_i\right)^T\\
\left({\bf B}_i{\bf V}^{(b)}_i\right)^T
\end{array}
\right]\hspace{-0.1cm}{\bf f}_i}_{\text{useful data}}\hspace{-0.1cm}+\hspace{-0.35cm}\sum_{j=1,j\ne i}^k\hspace{-0.1cm}
\underbrace{\left[
\begin{array}{@{}c@{}}
\left({\bf A}_{j}{\bf V}^{(a)}_i\right)^T\\
\left({\bf B}_j{\bf V}^{(b)}_i\right)^T\label{Yrep}
\end{array}
\right]\hspace{-0.1cm}{\bf f}_j}_{\text{interference by ${\bf f}_j$}}
\end{align}
}where ${\bf p}_i^{(a)},{\bf p}_i^{(b)}\in\mathbb{F}^{\frac{N}{2}}$ are the equations downloaded from parity nodes $a$ and $b$ respectively.
Here, ${\bf V}_i^{(a)},{\bf V}_i^{(b)}\in\mathbb{F}^{N\times \frac{N}{2}}$ denote the {\it repair matrices} used to mix the parity contents.\footnote{Here, we consider that the newcomer downloads the same amount of information from both parities. In general this does not need to be the case.}
Retrieving ${\bf f}_i$ from (\ref{Yrep}) is equivalent to solving an underdetermined set of $N$ equations in the $kN$ unknowns of ${\bf f}$, with respect to only the $N$ desired unknowns of ${\bf f}_i$.
However, this is not possible due to the additive {\it interference} components that corrupt the desired information in the received equations.
These terms are generated by the undesired unknowns ${\bf f}_j$, $j\ne i$, as noted in (\ref{Yrep}).
Additional data need to be downloaded from the systematic nodes, which will ``replicate'' the interference terms and will be subtracted from the downloaded equations. 
To erase a single interference term, a download of a basis of equations that generates the corresponding interference term, say $\left[
\begin{smallmatrix}
\left({\bf A}_{s}{\bf V}^{(a)}_i\right)^T\\
\left({\bf B}_s{\bf V}^{(b)}_i\right)^T\label{Yrep}
\end{smallmatrix}
\right]\hspace{-0.1cm}{\bf f}_j$, suffices.
Eventually, when all undesired terms are subtracted, a full rank system of $N$ equations in $N$ unknowns $\left[
\begin{smallmatrix}
\left({\bf A}_{i}{\bf V}^{(a)}_i\right)^T\\
\left({\bf B}_i{\bf V}^{(b)}_i\right)^T
\end{smallmatrix}
\right]{\bf f}_i$ has to be formed.
Thus, it can be proven that the {\it repair bandwidth} to exactly regenerate systematic node $i$ is given by
{\small
\begin{align}
\gamma_i=N+\sum_{j=1,j\ne i}^{k}\rank\left(\left[{\bf A}_j{\bf V}^{(a)}_i  \;{\bf B}_j{\bf V}^{(b)}_i\right]\right),\nonumber
\end{align}
}where the sum rank term is the aggregate of interference dimensions.
Interference alignment plays a key role since the lower the interference dimensions are, the less repair data need to be downloaded.
We would like to note that the theoretical minimum repair bandwidth of any node for optimal $(k+2,k)$ MDS codes is exactly $(k+1)\frac{N}{2}$, i.e. half of the remaining contents; this corresponds to each interference spaces having rank $\frac{N}{2}$.
This is also true for the systematic parts of non-MDS codes, as long as they have the same problem parameters that were discussed in the beginning of this section, and all the coding matrices have full rank $N$.
An abstract example of a code repair instance for  a $(4,2)$ storage code is given in Fig. 2, where interference terms are marked in red.

\begin{figure}[t]
\includegraphics[width=1\columnwidth]{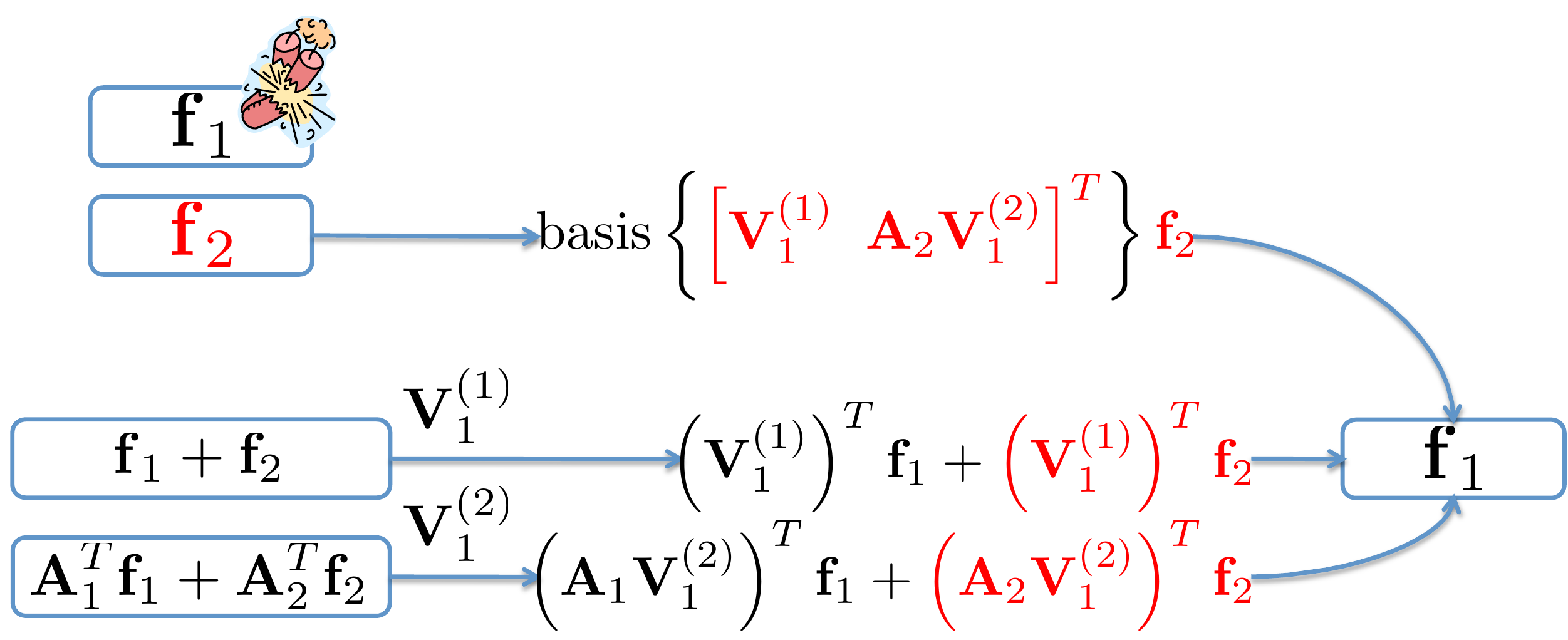}
\caption{Repair of a $(4,2)$ code.}
\end{figure}

To minimize the repair bandwidth $\gamma_i$, we need to carefully design both the storage code and the repair matrices.
In the following, we provide a $2$-parity code that achieves optimal systematic and near optimal parity repair.

\begin{figure*}[t!]
\begin{align}
{\bf X}_1 = \diag\left(\left[
\begin{smallmatrix}
 1\\
     1\\
     1\\
     1\\
    -1\\
    -1\\
    -1\\
    -1
\end{smallmatrix}
\right]
\right),
\;\;
{\bf X}_2 = \diag\left(
\left[
\begin{smallmatrix}
 1\\
     1\\
    -1\\
    -1\\
     1\\
     1\\
    -1\\
    -1
\end{smallmatrix}
\right]
\right),
\;\;
{\bf X}_3 = \diag\left(
\left[
\begin{smallmatrix}
 1\\
    -1\\
     1\\
    -1\\
     1\\
    -1\\
     1\\
    -1
\end{smallmatrix}
\right]
\right)\nonumber 
\end{align}
\hrulefill
\caption{The coding matrices of a repair optimal $(5,3)$ code over $\mathbb{GF}(3)$.}

\end{figure*}

\section{A New Storage Code}
We introduce a $(k+2,k)$ storage storage code over $\mathbb{GF}(3)$, for file sizes $M = k2^k$, with coding matrices
\begin{align}
{\bf A}_i &= {\bf I}_{N},\;\;  {\bf B}_i = {\bf X}_{i},\label{code}
\end{align}
where $N=2^k$, ${\bf X}_i = {\bf I}_{2^{i-1}}\otimes \text{blkdiag}\left({\bf I}_{\frac{N}{2^{i}}},-{\bf I}_{\frac{N}{2^{i}}}\right)$, and $i\in\{1,\ldots,k\}$.
In Fig. 3, we give the coding matrices of the $(5,3)$ version of the code.
\begin{theo}
The code in (\ref{code}) has optimally repairable systematic nodes and its parity nodes can be repaired by generating as much repair bandwidth as a systematic repair does. 
It can tolerate any single node failure, and any pair of failures that contains at most one systematic failure.
Moreover, to reconstruct the file at most $k+\frac{1}{2}$ coded blocks are required.
\end{theo}

In the following, we present the tools that we use in our derivations.
Then, in Sections V and VI we prove Theorem 1.

\section{Dots-on-a-Lattice and Hadamard Designs}
Optimality during a systematic repair, requires interference spaces collapsing  down to the minimum of $\frac{N}{2}$, out of the total $N$, dimensions.
At the same time, useful data equations have to span $N$ dimensions.
For the constructions presented here, we consider that the same repair matrix is used by both parities, i.e., ${\bf V}^{(1)}_i={\bf V}^{(2)}_i={\bf V}_i$.
Hence, for the repair of systematic node $i\in\{1,\ldots,k\}$ we optimally require 
\begin{equation}
\rank\left(\left[{\bf V}_i\; {\bf X}_{j}{\bf V}_i\right]\right)=\frac{N}{2},
\end{equation}
for all $j\in\{1,\ldots,k\}\backslash i$,
and at the same time
\begin{equation}
\rank\left(\left[{\bf V}_i\;\;{\bf X}_i{\bf V}_i\right]\right)=N.
\end{equation}
The key ingredient of our approach that eventually provides the above is Hadamard matrices.

To motivate our construction, we start by briefly discussing the repair properties of the asymptotic coding schemes of \cite{CadambeWinc}, \cite{SuhCodes}.
Consider a $2$-parity MDS storage code that requires file sizes $M = k2\Delta^{k-1}$, i.e., $N =2\Delta^{k-1}$. Its $N\times N$ diagonal coding matrices $\{{\bf X}_s\}_{s=1}^{k}$ have i.i.d. elements drawn uniformly at random from some arbitrarily large finite field $\mathbb{F}$.
During the repair of a systematic node $i\in\{1,\ldots,k\}$, the repair matrix ${\bf V}_i$ that is used by both parity nodes to mix their contents, has as columns the $\frac{N}{2}=\Delta^{k-1}$ elements of the set
{\small
\begin{equation}
\mathcal{V}_i=\left\{\prod_{s=1,s\ne i}^k{\bf X}_s^{x_s}{\bf w}: x_s\in\{0,\ldots,\Delta-1\}\right\}.
\end{equation}
}Then, we define a map $\mathcal{L}$ from vectors in the set $\left\{\prod_{s=1}^k{\bf X}_s^{x_s}{\bf w}:x_s\in\mathbb{Z}\right\}$ to points on the integer lattice $\mathbb{Z}^{k}$:
$\prod_{s=1}^{k}{\bf X}_s^{x_s}{\bf w} \overset{\mathcal{L}}{\rightarrow}  \sum_{s=1}^{k}x_s{\bf e}_s$,
where ${\bf e}_s$ is the $s$-th column of ${\bf I}_{k+1}$.
Now, consider the induced lattice representation of ${\bf V}_i$
{\small
\begin{equation}
\mathcal{L}({\bf V}_i)\eqdef\left\{\sum_{s=1,s\ne i}^kx_s{\bf e}_s;\; x_s\in\{0,\ldots,\Delta-1\}\right\}.
\end{equation}
}Observe that the $i$-th dimension of the lattice where $\mathcal{L}({\bf V}_i)$ lies on, indicates all possible exponents $x_i$ of ${\bf X}_i$.
\begin{figure}[t]
\begin{center}
 \includegraphics[width=0.7\columnwidth]{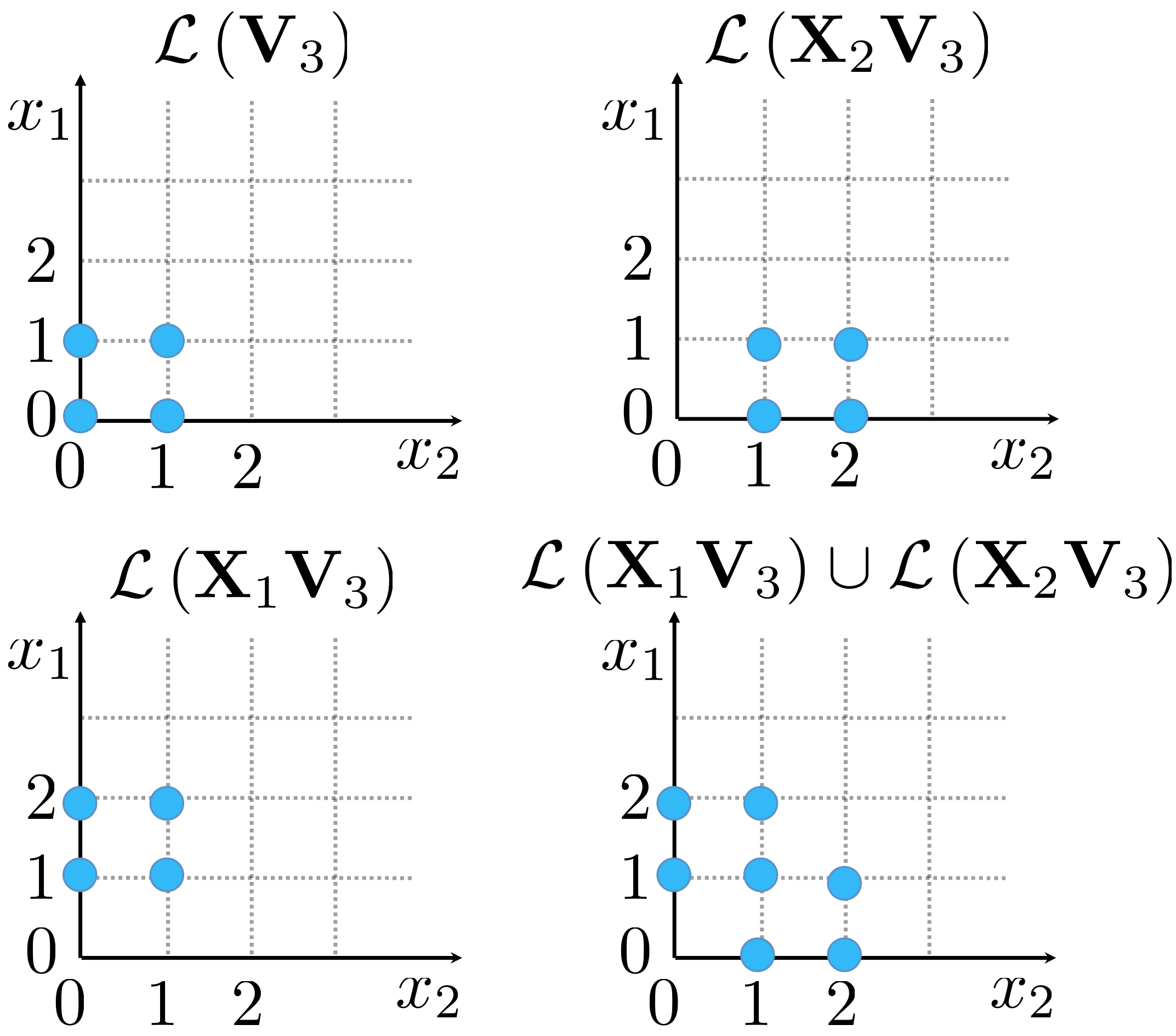}
\caption{Here we have $k=3$, $\frac{N}{2}=4$, and $\Delta=2$. Moreover, $\mathcal{L}({\bf V}_3)=\left\{(0,0,0),(0,1,0),(1,0,0),(1,1,0)\right\}$, $\mathcal{L}({\bf X}_1{\bf V}_3)=\left\{(1,0,0),(1,1,0),(2,0,0),(2,1,0)\right\}$, and $\mathcal{L}({\bf X}_2{\bf V}_3)=\left\{(0,1,0),(0,2,0),(1,1,0),(1,2,0)\right\}$.}
\end{center}
\end{figure}
Then, the products ${\bf X}_j{\bf V}_i$, $j\ne i$, and ${\bf X}_i{\bf V}_i$ map to 
{\small
\begin{align}
\mathcal{L}({\bf X}_j{\bf V}_i)&=\Biggl\{\hspace{-0.1cm}(x_j+1){\bf e}_j\hspace{-0.1cm}+\hspace{-0.3cm}\sum_{s=1,s\ne j}^k\hspace{-0.3cm}x_s{\bf e}_s;\; x_s\in\{0,\ldots,\Delta-1\}\Biggr\}\nonumber\\
\text{and }\mathcal{L}({\bf X}_i{\bf V}_i)&=\Biggl\{e_i+\sum_{i=1,s\ne i}^kx_i{\bf e}_i;\; x_s\in\{0,\ldots,\Delta-1\}\Biggr\},\nonumber
\end{align}
}respectively.
In Fig. 2, we give an illustrative example for $k=3$, and $\Delta=2$.
\begin{rem}
Observe how matrix multiplication of ${\bf X}_i$ and elements of $\mathcal{V}_i$ manifests itself through the dots-on-a-lattice representation: the product of ${\bf X}_i$ with the  elements of $\mathcal{V}_i$ shifts the corresponding arrangement of dots along the $x_i$-axis, i.e., the $x_i$-coordinate of the initial points gets increased by one.
\end{rem}

Asymptotically optimal repair of node $i$ is possible due to the fact that interference spaces asymptotically align
{\small
\begin{align}
\frac{\rank\left(\left[{\bf V}_i\;\;{\bf X}_j{\bf V}_i\right]\right)}{\frac{N}{2}}&=\frac{\left|\mathcal{L}({\bf V}_i)\cup\mathcal{L}({\bf X}_j{\bf V}_i)\right|}{{\Delta^{k-1}}}\nonumber\\
& = \frac{\left|\mathcal{L}({\bf V}_i)\right|+o(\Delta^{k-1})}{{\Delta^{k-1}}} \overset{\Delta\rightarrow\infty}{\longrightarrow} 1,
\end{align}
}and useful spaces span $N$ dimensions, that is,
$\rank\left(\left[{\bf V}_i\;\;{\bf X}_i{\bf V}_i\right]\right) = \left|\mathcal{L}({\bf V}_i)\cup\mathcal{L}({\bf X}_i{\bf V}_i)\right| = 2\Delta^{k-1}$,
with arbitrarily high probability for sufficiently large field sizes.

The question that we answer here is the following: How can we design the coding and the repair matrices such that {\it i)} {\it exact} interference alignment is possible and {\it ii)} the full rank property is satisfied, for fixed in $k$ file size and field size?
We first address the first part. 
We want to design the code such that the space of the repair matrix is invariant to any transformation by matrices generating its columns, i.e., 
 $\mathcal{L}({\bf X}_j{\bf V}_i)=\mathcal{L}({\bf V}_i)$. This is possible when
{\footnotesize
\begin{align}
\mathcal{L}({\bf X}_j{\bf V}_i)&=\Biggl\{\hspace{-0.1cm}(x_j+1){\bf e}_j\hspace{-0.1cm}+\hspace{-0.3cm}\sum_{s=1,s\ne j}^k\hspace{-0.3cm}x_s{\bf e}_s;\; x_s\in\{0,\ldots,\Delta-1\}\Biggr\}\nonumber\\
&=\Biggl\{\hspace{-0.1cm}x_j{\bf e}_j\hspace{-0.1cm}+\hspace{-0.3cm}\sum_{s=1,s\ne j}^k\hspace{-0.3cm}x_s{\bf e}_s;\; x_s\in\{0,\ldots,\Delta-1\}\Biggr\}=\mathcal{L}({\bf V}_i),\nonumber
\end{align}
}that is,  when the matrix powers ``wrap around'' upon reaching their modulus $\Delta$.
This wrap-around property is obtained when the diagonal coding matrices have elements that are roots of unity. 
\begin{lem}
For diagonal matrices, ${\bf X}_1,\ldots,{\bf X}_k$, whose elements are $\Delta$-th roots of unity, i.e., ${\bf X}_s^{\Delta} = {\bf X}_s^0$, for all $s\in\{1,\ldots,k\}$, we have that
$\mathcal{L}({\bf X}_j{\bf V}_i)=\mathcal{L}({\bf V}_i)$, for all $i\in\{1,\ldots,k\}\backslash j$.
\end{lem}

However, arbitrary diagonal matrices whose elements are roots of unity are not sufficient to ensure the full rank property of the useful data repair space $\left[{\bf V}_i\;\;{\bf X}_i{\bf V}_i\right]$.
In the following we prove that the full rank property along with perfect IA is guaranteed
when we set $N=2^k$, ${\bf X}_i = {\bf I}_{2^{i-1}}\otimes \text{blkdiag}\left({\bf I}_{\frac{N}{2^{i}}},-{\bf I}_{\frac{N}{2^{i}}}\right)$, and consider the set
\begin{equation}
\mathcal{H}_{N} = \left\{\prod_{i = 1}^{k}{\bf X}_{i}^{x_i}{\bf w}: x_{i}\in \{0,1\}\right\}. \label{Hprod}
\end{equation}
Interestingly, there is a one-to-one correspondence between the elements of $\mathcal{H}_{N}$ and the columns of a Hadamard matrix.
\begin{lem}
Let  an $N\times N$ Hadamard matrix of the Sylvester's construction
{\small
\begin{equation}
{\bf H}_{N} \eqdef \left[
\begin{array}{rr}
{\bf H}_{\frac{N}{2}} &{\bf H}_{\frac{N}{2}}\\
{\bf H}_{\frac{N}{2}} & -{\bf H}_{\frac{N}{2}}
\end{array}
\right],
\end{equation}
}with ${\bf H}_{1} = 1$.
Then,
${\bf H}_{N}$ is full-rank with mutually orthogonal columns, that are the $N$ elements of $\mathcal{H}_{N}$.
Moreover, any two columns of ${\bf H}_{N}$ differ in $\frac{N}{2}$ positions.
\label{HadamardLem}
\end{lem}
The proof is omitted due to lack of space.
To illustrate the connection between $\mathcal{H}_{N}$ and ${\bf H}_{N}$ we ``decompose'' the Hadamard matrix of order $4$
{\small
\begin{align}
{\bf H}_4 &= \left[
\begin{smallmatrix}
1 & 1 & 1 & 1\\
1 & -1 & 1 & -1\\
1 & 1 & -1 & -1\\
1 & -1 & -1 & 1
\end{smallmatrix}
\right] = \left[{\bf w}\;\;{\bf X}_2{\bf w}\;\;{\bf X}_1{\bf w}\;\;{\bf X}_2{\bf X}_1{\bf w}\right],
\end{align}
}where 
${\bf X}_1 = \text{diag}\left(
\begin{smallmatrix}
1\\
1\\
-1\\
-1
\end{smallmatrix}
\right) \text{ and } 
{\bf X}_2 = 
\diag\left(
\begin{smallmatrix}
1 \\
-1\\
1\\
-1
\end{smallmatrix}
\right)$.
Due to the commutativity of ${\bf X}_1$ and ${\bf X}_2$, the columns of ${\bf H}_{4}$ are also the elements of
$\mathcal{H}_{4}=\left\{{\bf w},{\bf X}_1{\bf w},{\bf X}_2{\bf w},{\bf X}_1{\bf X}_2{\bf w}\right\}$.

By using $\mathcal{H}_N$ as our ``base'' set, we are able to obtain perfect alignment condition due to the wrap around property of it elements; the full rank condition will be also satisfied due to the mutual orthogonality of these elements.


\begin{figure}[t]
\begin{center}
\includegraphics[width=0.7\columnwidth]{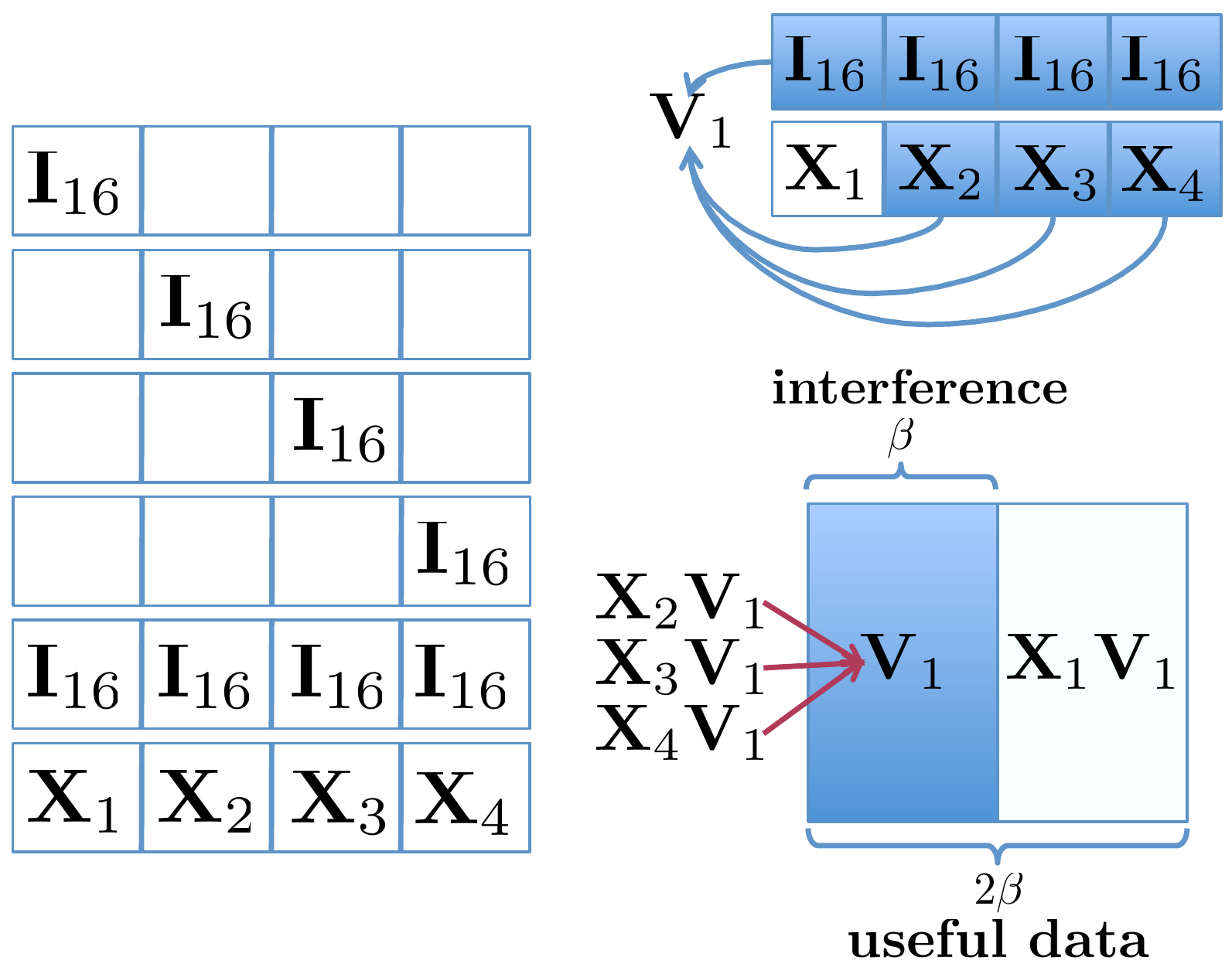}
\end{center}
\caption{The coding matrices of our $(6,4)$ code are given. We illustrate the ``absorbing'' properties of the repair matrix for systematic node $1$. The column space of the repair matrices is invariant to the corresponding blue blocks. This results in interference spaces aligning in exactly half of the dimensions available.}
\end{figure}

\section{Repairing Single Node Failures}
\subsection{Systematic Repairs}
Let systematic node $i\in\{1,\ldots,k\}$ fail.
Then, we pick the columns of the repair matrix as a set of $\frac{N}{2}$ vectors whose lattice representation is invariant to all ${\bf X}_j$s  but to one key matrix ${\bf X}_i$.
We specifically construct the $N\times\frac{N}{2}$ repair matrix ${\bf V}_i$ whose columns have a one-to-one correspondence with the elements of the set
\begin{equation}
\mathcal{V}_i = \left\{\prod_{s=1,s\ne i}^{k} {\bf X}^{x_s}_s{\bf w}:x_s\in\{0,1\}\right\}. \label{Vi}
\end{equation}
First, observe that ${\bf V}_i$  is full column rank since it is a collection of $\frac{N}{2}$ distinct columns from $\mathcal{H}_{N}$.
Then, we have the following lemma.
\begin{lem}
For any $i,j\in\{1,2,\ldots,k\}$, we have that
\begin{align}
\rank(\left[{\bf V}_i \;{\bf X}_j{\bf V}_i\right])&=\left|\mathcal{L}({\bf V}_i)\cup\mathcal{L}\left({\bf X}_j{\bf V}_i\right)\right|\nonumber\\
&=\left\{
\begin{array}{lc}
N, & i=j\\
\frac{N}{2}, & i \ne j
\end{array}
\right..
\end{align}
\end{lem}
The above holds due to each element of $\mathcal{H}_N$ being associated with a unique power tuple.
Then, the columns of $\left[{\bf V}_i \;{\bf X}_i{\bf V}_i\right]$ are exactly the elements of $\mathcal{H}_N$, since
{\footnotesize
\begin{equation}
\begin{split}
\mathcal{L}\left({\bf V}_i\right)\cup\mathcal{L}\left({\bf X}_{i}{\bf V}_i\right) &=\left\{\sum_{s=1,s\ne i}^{k}x_i{\bf e}_i;\; x_i\in\{0,1\}\right\}\\
&\bigcup \left\{e_i+\sum_{s=1,s\ne i}^{k}x_i{\bf e}_i;\; x_i\in\{0,1\}\right\}\\
&=\mathcal{L}\left({\bf H}_{N}\right).
\end{split}
\end{equation}
}Moreover, the set of columns in ${\bf V}_i$ are identical to the set of columns of ${\bf X}_j{\bf V}_i$, i.e., $\mathcal{L}({\bf V}_i)=\mathcal{L}({\bf X}_j{\bf V}_i)$, for $j\ne i$, due to Lemmata 1 and 2.
Therefore, the interference spaces span $\frac{N}{2}$ dimensions, which is the theoretic minimum,
and the desired data space during any systematic node repair is full-rank, since it has as columns all columns of ${\bf H}_N$.

Hence, we conclude that a single systematic node of the code can be repaired with bandwidth
$(k+1)\frac{N}{2}=\frac{k+1}{2k}M$.
In Fig. 4, we depict a $(6,4)$ code of our construction, along with the illustration of the repair spaces.

\subsection{Parity repairs}
Here, we prove that a single parity node repair generates at most the repair bandwidth of a single systematic repair.
Let parity node $a$ fail. 
Then, observe that if the newcomer uses the $N\times N$ repair matrix ${\bf V}_a^{(b)}={\bf X}_1$ to multiply the contents of parity node $b$,
then it downloads
${\bf X}_1\left(\sum_{i=1}^k{\bf X}_1{\bf f}_i\right)={\bf f}_1+\sum_{i=2}^k{\bf X}_1{\bf X}_i{\bf f}_i$.
Observe, that the component corresponding to systematic part ${\bf f}_1$ appears the same in the linear combination stored at the lost parity. 
By Lemma 2, each of the remaining blocks, ${\bf X}_1{\bf X}_i{\bf f}_i$ share exactly $\frac{N}{2}$ indices with equal elements to the same $\frac{N}{2}$ indices of ${\bf X}_i{\bf f}_i$ which was lost, for any $i\in\{2,\ldots,k\}$.
This is due to the fact that the diagonal elements of matrices ${\bf X}_1{\bf X}_i$ and ${\bf X}_i$ are the elements of some two columns of ${\bf H}_{N}$.
Therefore, the newcomer has to download from systematic node $j\in\{2,\ldots,k\}$, the $\frac{N}{2}$ entries that parity $a$'s component ${\bf X}_j{\bf f}_j$ differs from the term ${\bf X}_1{\bf X}_j{\bf f}_j$ of the downloaded linear combination. 
Hence, the first parity can be repaired with bandwidth at most $N+(k-1)\frac{N}{2}=(k+1)\frac{N}{2}$.\footnote{By ``at most'' we mean that this result is proved using an achievable scheme, however, we do not prove that it is optimal.}
The repair of parity node $b$ can be performed in the same manner.

\section{Erasure Resiliency}
Our code can tolerate any single node failure and any two failures with at most one of them being a systematic one.
A double systematic and parity node failure can be treated by first reconstructing the lost systematic node from the remaining parity, and then reconstructing the lost parity from all the systematic nodes.
However, two simultaneous systematic node failures cannot be tolerated.
Consider for example the corresponding matrix when we connect to nodes $\{1,\ldots,k-2\}$ and both parities:
{\scriptsize
\begin{equation}
\left[
\begin{array}{ccc|cc}
{\bf I}_{N} & \ldots & {\bf 0}_{N\times N}&{\bf 0}_{N\times N}&{\bf 0}_{N \times N}\\
\vdots&&\vdots&\vdots\\
{\bf 0}_{N\times N} &\ldots &  {\bf I}_{N}&{\bf 0}_{N \times N}&{\bf 0}_{N \times N }\\
\hline
{\bf I}_{N} &\ldots&{\bf I}_{N}&{\bf I}_{N}&{\bf I}_{N}\\
{\bf X}_1 &\ldots&{\bf X}_{k-2}&{\bf X}_{k-1}&{\bf X}_k
\end{array}
\right]{\bf f}.
\label{DC_2}
\end{equation}
}The rank of this $kN\times kN$ matrix is $(k-1)N+\frac{N}{2}$ due to the submatrix
$\left[
\begin{smallmatrix}
{\bf I}_{N}&{\bf I}_{N}\\
{\bf X}_{k-1}&{\bf X}_k
\end{smallmatrix}
\right]$ having rank $\frac{3N}{2}$.
For these cases, an extra download of $\frac{N}{2}$ equations is required to decode the file, i.e., an aggregate download of $kN+\frac{N}{2}$ equations, or $k+\frac{1}{2}$ encoded pieces.

\end{document}